# TOMOGRAPHY OF MULTI-PHOTON POLARIZATION STATES IN CONDITIONS OF NON-UNIT QUANTUM EFFICIENCY OF DETECTORS

Yu. I. Bogdanov*[1,2,3], B. I. Bantysh**[1,3], N. A. Bogdanova[1,3], V. F. Lukichev[1]

1 – Institute of Physics and Technology, Russian Academy of Sciences, 117218, Moscow, Russia
2 – National Research Nuclear University "MEPHI", 115409, Moscow, Russia
3 – National Research University of Electronic Technology MIET, 124498, Moscow, Russia

The polarizing multi-photon quantum states tomography with non-unit quantum efficiency of detectors is considered. A new quantum tomography protocol is proposed. This protocol considers events of losing photons of multi-photon quantum state in one or more channels among with n-fold coincidence events. The advantage of the proposed protocol compared with the standard n-fold coincidence protocol is demonstrated using the methods of statistical analysis.

## INTRODUCTION

Quantum states of light are one of the most prospective physical resources for quantum computation realization. In particular, polarizing quantum states of light are already being used for quantum cryptography [1,2] and are suitable for one-way quantum computations [3,4]. The most important tool for the control of quantum technologies is provided by quantum states tomography. Because of the statistical nature of quantum mechanics and N. Bohr's complementarity principle [5,6] one should perform a set of complementary measurements to obtain enough information about the quantum state. Next, using the obtained data the procedure of quantum state reconstruction is performed. For this purpose, in the current work we use the root approach [7-10] together with the maximum likelihood method [11-15].

While collecting the statistical data in real experiments one faces various kinds of noises and errors which could come out at any stage of the quantum tomography procedure. In particular, for polarizing quantum states the non-unit detectors quantum efficiency is a severe source of errors. In typical laboratory conditions quantum efficiency is to the order of 20% [16]. Thereby, for multi-photon polarizing states tomography the events of photon registration in each channel (n-fold coincidence) become very rare. Today this problem is especially topical because the frequency of generation of entangled multi-photon polarization quantum states in present experiments is very low, for instance, for three-photons states it is under 1 Hz [16].

The approach proposed in the current work is based on the formalism of fuzzy measurements [17] and lets us construct the protocol of multi-photon quantum states tomography that extracts information from the events of 1-fold, 2-fold, etc. coincidence among with n-fold coincidence events. As it will be shown, this approach enables one to significantly increase the accuracy of quantum tomography of multi-photon polarizing quantum states with a fixed sample size.

* bogdanov_yurii@inbox.ru
** bbantysh60000@gmail.com

## POLARIZATION QUANTUM STATE TOMOGRAPHY

Let us consider the measurement of an arbitrary pure *N*-photon polarizing quantum state $|\psi\rangle$. One should notice that all the reasoning below could be also extended to mixed states by using the state purification procedure [18]. The state $|\psi\rangle$ is determined by $2s - 2$ independent real parameters, where $s = 2^N$ is the dimension of the Hilbert space of this state. To determine these parameters one should prepare an ensemble of *n* identical *N*-photon quantum states $|\psi\rangle$ and perform a set of complementary projective measurements over photons in each of *N* channels.

Projective measurement of unknown polarizing states of photon (Figure 1) is based on the use of a polarizing beam splitter (PBS), two detectors (D1 and D2), one half-wave plate (HWP) and one quarter-wave plate (QWP) [19]. By varying the angles of plates rotation one can set the required unitary transformation *U*. The considered scheme provides the simultaneous measurement of two orthogonal projectors $U^+|H\rangle\langle H|U$ and $U^+|V\rangle\langle V|U$ for the input state $|\psi\rangle$ where $|H\rangle$ and $|V\rangle$ denote the states of photon with horizontal and vertical polarizations respectively. We consider the quantum efficiencies of detectors D1 and D2 to be equal to $\eta_1 = \eta_2 = \eta$.

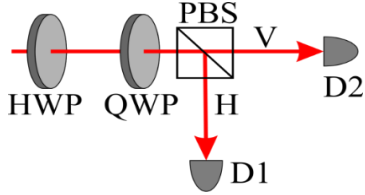

*Figure 1. The scheme of projective measurement for polarizing quantum state [19]; HWP and QWP are half-wave plate and quarter-wave plate, PBS is the polarizing beam splitter, D1 and D2 are photon detectors*

Let the measurement protocol for each of *N* channels consist of $m_1$ projectors $\Lambda_{1,j}$ ( $j = 1, 2, \ldots m_1$ ) each. The full protocol with $m = m_1^N$ projectors $\Lambda_j$ ( $j = 1, 2, \ldots m$ ) is formed by all the possible tensor products of one-photon projectors. This corresponds to the projection of *N*-qubit quantum state $|\psi\rangle$ to unentangled states.

According to the Born rule the probability of event registration per unit of time for each projector is defined:

$$\lambda_j(\psi) = \langle\psi|\Lambda_j|\psi\rangle, \qquad j = 1, 2, \ldots m. \tag{1}$$

During the measurement process for each of *n* representatives of quantum state $|\psi\rangle$ the phase plates configuration is selected to fit one of *m* projectors. We will consider the case when for each configuration an equal amount of state representatives is allocated. Then the exposition time, which sets the weight for each projector, is the same and is equal to

$$t_j = \frac{ns}{m}, \qquad j = 1, 2, \ldots m. \tag{2}$$

We consider protocols that form the unity decomposition. In our case weights (2) are set to make the sum of all protocol matrices $t_j \Lambda_j$ proportional to the identity matrix:

$$\sum_{j=1}^{m} t_j \Lambda_j = n I_s \qquad (3)$$

Here $I_s$ is the identity matrix of dimension $s \times s$. Let us notice that the condition (3) together with equation (1) and vector-state normalization condition $\langle \psi | \psi \rangle = 1$ directly result in the full expected number of counts to be normalized to the full sample size ($n$ is a full sample size by all the elements of protocol):

$$\sum_{j=1}^{m} t_j \lambda_j = n. \qquad (4)$$

The measurements result is $m$ integers $k_j$ ($j = 1, 2, \ldots m$) corresponding to the number of events for each projector. The number of registered events $k_j$ has the Poisson distribution with the mean value $\lambda_j t_j$.

The problem of the state $|\psi\rangle$ reconstruction is reduced to the mathematical procedure of finding the quantum state, which gives probabilities (1) that fit experimental counts $k_j$ in the best way. The maximum likelihood method implemented within the root approach gives the following convergent quasi-linear equation [7-9]:

$$I |\psi\rangle = J(\psi) |\psi\rangle, \qquad I = \sum_{j=1}^{m} t_j \Lambda_j, \qquad J(\psi) = \sum_{j=1}^{m} \frac{k_j}{\lambda_j(\psi)} \Lambda_j. \qquad (5)$$

The numerical solution of this equation lets us estimate the state $|\psi\rangle$ by the set of registered events $k_j$.

## DETECTORS IMPERFECTION

Let us consider the non-unit quantum efficiency as a type of detectors non-ideality. This means that with a certain probability the photon is not registered by the detector. Thus to perform the $N$-photon state tomography one usually resorts to n-fold coincidence scheme [20]. This scheme considers only completely successful events when photons are registered in each of $N$ channels. However, this approach has a significant drawback as it discards useful information contained in other events of losing only one, two, etc. photons. This information could be taken into consideration using the method of fuzzy measurements [17]. This method is the generalization of ideal measurements described above.

We will perform the formal description of fuzzy measurements that takes into account detectors non-unit quantum efficiency using the scheme on Figure 1 in the following way. For each channel we introduce $m_1 + 1$ measurement operators: $m_1$ projectors for successful measurement events and one additional operator $I_2/2$ that describes completely chaotic state and corresponds to unregistered photon. The full protocol for $N$ channels consists of $m = (m_1 + 1)^N$ measurement operators and includes events of losing any number of photons.

Taking non-unit quantum efficiency into account one should also specify weights $t_j$. This weights differ from those for the ideal protocol (2):

$$t_{j \in J_k} = \frac{ns}{m_1^k} \eta^k (1-\eta)^{(N-k)}, \qquad k = 0, 1, \ldots, N. \tag{6}$$

Symbol $J_k$ denotes the set of all the indices of protocol elements which correspond to photons registration in $k$ channels exactly (in all the other $N - k$ channels photons aren't registered). The weights in (6) are set in the way that keeps the measurement protocol to form the unity decomposition (meet condition (3)).

To estimate the efficiency of the proposed protocol it is convenient to use the complete information matrix which is the generalization of the Fisher information matrix [18]:

$$H = 2 \sum_{j=1}^{m} \frac{t_j}{\lambda_j} \left( \tilde{\Lambda}_j | \tilde{\psi} \rangle \right) \left( \tilde{\Lambda}_j | \tilde{\psi} \rangle \right)^+, \tag{7}$$

where real measurement operators and quantum state under consideration are introduced:

$$\tilde{\Lambda}_j = \begin{pmatrix} \operatorname{Re}(\Lambda_j) & -\operatorname{Im}(\Lambda_j) \\ \operatorname{Im}(\Lambda_j) & \operatorname{Re}(\Lambda_j) \end{pmatrix}, \quad |\tilde{\psi}\rangle = \begin{pmatrix} \operatorname{Re}(|\psi\rangle) \\ \operatorname{Im}(|\psi\rangle) \end{pmatrix}.$$

These values correspond to transition between the initial Hilbert space and the real Euclidean space of a doubled dimension.

The eigenvalues of matrix (7) let us evaluate the statistical fluctuations of the quantum state parameters under reconstruction. For the state reconstruction accuracy estimation we use the standard measure of the overlap between the true state $|\psi\rangle$ and the reconstructed one $|\psi_{rec}\rangle$: fidelity $F = |\langle \psi | \psi_{rec} \rangle|^2$.

## NUMERICAL RESULTS

Figure 2 depicts theoretical distributions of fidelity for the reconstruction of three-photon quantum state GHZ ($N = 3$) and for different values of detectors quantum efficiency. Theoretical distributions were obtained by analyzing the matrix (7). As the measure of fidelity we have considered parameter $z = -\log_{10}(1-F)$. This parameter characterizes the number of nines in the decimal representation of the fidelity $F$. For example, $z = 3$ corresponds to $F = 0,999$. As the set of one-photon projectors here we have used the protocol with octahedron symmetry ($m_1 = 8$) [21]. The sample size has been set to $n = 10^5$. For quantum efficiency $\eta = 0,2$ the protocol which has been described above and is based on fuzzy measurements gives the average fidelity losses $\langle 1-F \rangle_{fuzzy} \approx 0,00274$ while the n-fold coincidence protocol gives $\langle 1-F \rangle_{coinc} \approx 0,00967$ that is approximately 3,5 times higher. For the values of quantum efficiency $\eta = 0,4$ and $\eta = 0,6$ we have the gain in the accuracy of the developed protocol by approximately 2,2 and 1,6 times respectively. In the ideal case of unit quantum efficiency of detectors both protocols are completely equivalent and give average fidelity losses $\langle 1-F \rangle \approx 0,0000774$.

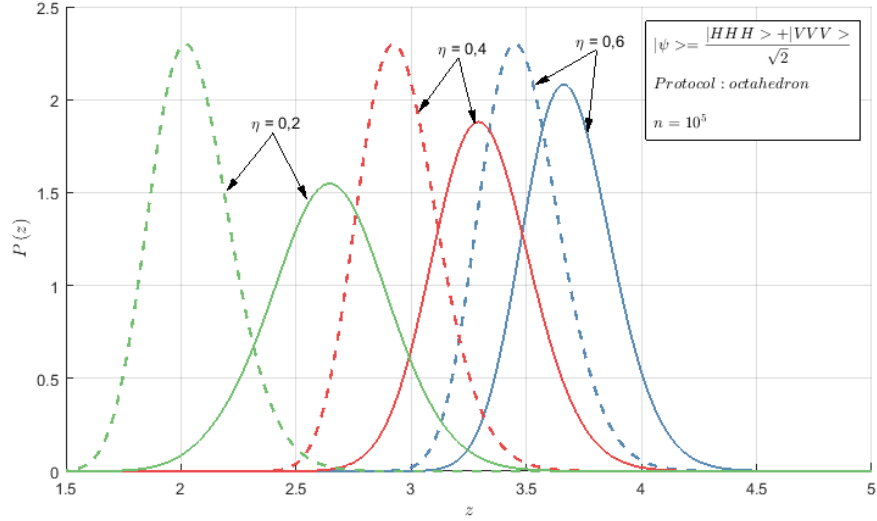

*Figure 2. The distribution of parameter z for the state GHZ reconstruction; octahedron symmetry protocol; sample size – $10^5$; solid lines denote the distribution for protocol based on fuzzy measurements, dashed lines denote the protocol, based on n-fold coincidence*

Figure 3 depicts the results of numerical experiments of the state GHZ reconstruction by using the maximum likelihood method (5) with the protocol based on fuzzy measurements. The detectors efficiency has been set to $\eta = 0,6$. In each of 200 experiments which have been carried out the sample size has been set to $n = 10^5$. As the set of one-photon projectors we have used the protocol with octahedron symmetry ($m_1 = 8$). Also the figure depicts the theoretical distribution of the parameter $z$ which has been obtained using the complete information matrix (7). The chi-squared criteria demonstrates that the developed theory agrees with the results of numerical experiments.

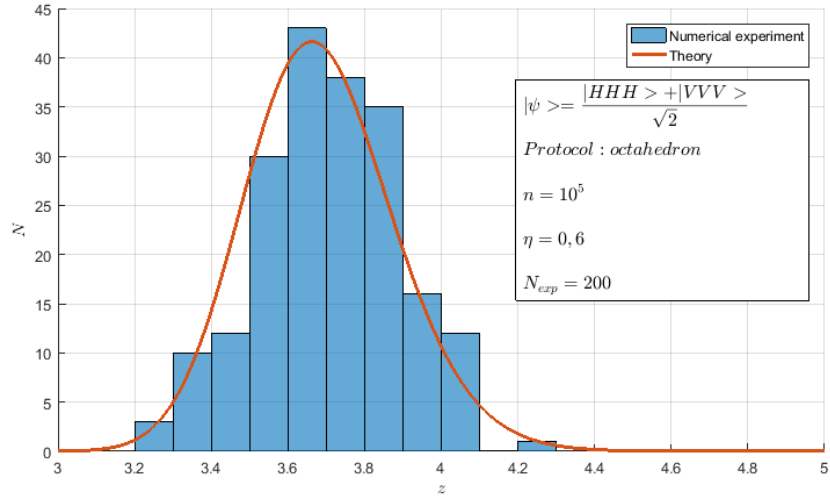

*Figure 3. The distribution of parameter z for the state GHZ reconstruction with the protocol based on fuzzy measurements (numerical experiments); solid line denotes the theoretical distribution; detectors efficiency – 0,6; octahedron symmetry protocol; number of experiments – 200; sample size in each experiment – $10^5$*

Let us note that the theory of fuzzy measurements fidelity developed in the current research generalizes the theory of quantum measurements fidelity developed within our previous works [18, 21].

One could estimate the degree of detectors non-ideality influence by analyzing the full information gathered by measurements. The full information is defined as a trace of information matrix (7). Information matrix is proportional to the sample size and its trace is equal to $2ns$ in the case of ideal measurements. Thereby, for convenience we consider the full information normalized to $2ns$:

$$h = \frac{1}{2ns}\text{Tr}(H) = \frac{1}{ns}\sum_{j=1}^{m} t_j \frac{\langle\psi|\Lambda_j^2|\psi\rangle}{\lambda_j}. \qquad (8)$$

For any ideal projective measurements $\Lambda_j^2 = \Lambda_j$ so

$$h_{ideal} = \frac{1}{ns}\sum_{j=1}^{m} t_j = 1.$$

Let us note that for non-ideal measurements $h<1$. In the case of detectors non-unit quantum efficiency the n-fold coincidence protocol differs from ideal one in the reduced sample size. For this protocol the normalized full information is

$$h_{coinc} = \eta^N.$$

For the fuzzy measurements protocol considered in the current work

$$h_{fuzzy} = \left(\frac{1+\eta}{2}\right)^N,$$

Thus, information $h_{fuzzy}$ which is contained in fuzzy measurements is higher than the information $h_{coinc}$ for the n-fold coincidence protocol by $\frac{1}{s}\left(1+\frac{1}{\eta}\right)^N$ times. Let us note that the advantage in information for the developed protocol sharply increases with the increase in the number of channels $N$ and the decrease in detectors quantum efficiency $\eta$.

## SUMMARY

Let us overview the main results of this work.

The protocol of multi-photon polarizing quantum state tomography that takes non-unit detectors quantum efficiency into consideration has been proposed.

Using the generalization of Fisher information matrix the information characteristics of quantum measurements for three different protocols have been investigated: fuzzy measurements protocol proposed in the current work, n-fold coincidence protocol and ideal protocol.

By means of analytical consideration and numerical statistical simulation the comparison between the proposed fuzzy measurements protocol and the standard n-fold coincidence protocol has been made. It has been shown that the proposed protocol lets us reconstruct quantum states with much higher accuracy in comparison to the standard protocol.

## ACKNOWLEDGEMENTS

The work was supported by the Russian Science Foundation, grant № 14-12-01338 П.